\documentclass[twocolumn,aps,prl]{revtex4-1}
\usepackage{lineno}

\usepackage[percent]{overpic}
\usepackage{etoolbox}

\usepackage{graphicx,framed,amssymb,mathrsfs,textcomp,comment,fancyhdr,calligra,ifsym,amsmath,amsthm,mathtools,amsmath,float}

\usepackage[T1]{fontenc}
\usepackage[utf8]{inputenc}
\begin{document}
\title{{\bfseries A Storage Ring Experiment to Detect a Proton Electric Dipole Moment}}
\author{V. Anastassopoulos$^{16}$, S. Andrianov$^{30}$, R. Baartman$^{25}$, M. Bai$^{8}$, S. Baessler$^{20}$, J. Benante$^{2}$, M. Berz$^{15}$, M. Blaskiewicz$^{2}$, T. Bowcock$^{27}$, K. Brown$^{2}$,  B. Casey$^{26}$, M. Conte$^{31}$, J. Crnkovic$^{2}$, G. Fanourakis$^{5}$, A. Fedotov$^{2}$, P. Fierlinger$^{29}$, W. Fischer$^{2}$, M.O. Gaisser$^{23}$, Y. Giomataris$^{19}$,  M. Grosse-Perdekamp$^{10}$,
G. Guidoboni$^{7}$, S. Hac{\i}\"{o}mero\u{g}lu$^{23}$, G. Hoffstaetter$^{4}$, H. Huang$^{2}$, M. Incagli$^{17}$, A. Ivanov$^{30}$, D. Kawall$^{14}$,  B. Khazin$^{3,\dagger}$, Y.I. Kim$^{23}$, B. King$^{27}$, I.A. Koop$^{3}$, R. Larsen$^{2}$, D.M. Lazarus$^{2}$, V. Lebedev$^{26}$, M.J. Lee$^{23}$, S. Lee$^{23}$, Y.H. Lee$^{28}$, A. Lehrach$^{8}$, 
P. Lenisa$^{7}$, P. Levi Sandri$^{9}$, A.U. Luccio$^{2}$,
A. Lyapin$^{13}$, W. MacKay$^{2}$, R. Maier$^{8}$, K. Makino$^{15}$, N. Malitsky$^{2}$, W.J. Marciano$^{2}$, W. Meng$^{2}$, F. Meot$^{2}$,
E.M. Metodiev$^{22,23}$, L. Miceli$^{23}$, D. Moricciani$^{18}$, W.M. Morse$^{2}$, S. Nagaitsev$^{26}$, S.K. Nayak$^{2}$,  Y.F. Orlov$^{4}$,
C.S. Ozben$^{12}$, S.T. Park$^{23}$, A. Pesce$^{7}$, P. Pile$^{2}$, V. Polychronakos$^{2}$, B. Podobedov$^{2}$, J. Pretz$^{21}$, V. Ptitsyn$^{2}$, E. Ramberg$^{26}$,  D. Raparia$^{2}$,
 F. Rathmann$^{8}$, S. Rescia$^{2}$, T. Roser$^{2}$, H. Kamal Sayed$^{2}$, Y.K. Semertzidis$^{23,24,*}$, Y. Senichev$^{8}$, A. Sidorin$^{6}$, A. Silenko$^{1,6}$,
 N. Simos$^{2}$, A. Stahl$^{21}$, E.J. Stephenson$^{11}$, H. Str\"oher$^{8}$, M.J. Syphers$^{15}$, J. Talman$^{2}$, R.M. Talman$^{4}$,  V. Tishchenko$^{2}$, C. Touramanis$^{27}$, N. Tsoupas$^{2}$, G. Venanzoni$^{9}$, K. Vetter$^{2}$, S. Vlassis$^{16}$, E. Won$^{23,32}$, G. Zavattini$^{7}$, A. Zelenski$^{2}$, K. Zioutas$^{16}$\\ (Storage Ring EDM Collaboration)}
\affiliation{
$^{1}$Research Inst. for Nucl. Probl. of Belarusian State University, Minsk, Belarus \\
$^{2}$Brookhaven National Laboratory, Upton, NY 11973, USA \\
$^{3}$Budker Institute of Nuclear Physics, 630090 Novosibirsk, Russia\\
$^{4}$Laboratory for Elementary-Particle Physics, Cornell University, Ithaca, NY 14853, USA\\ 
$^{5}$Inst. of Nuclear Physics NCSR Demokritos, GR-15310 Aghia Paraskevi Athens, Greece \\
$^{6}$Joint Institute for Nuclear Research, Dubna, Moscow region, Russia \\
$^{7}$University of Ferrara, INFN of Ferrara, Ferrara, Italy \\
$^{8}$Institut f\"ur Kernphysik and JARA-Fame, Forschungszentrum J\"ulich, 52425 J\"ulich, Germany \\
$^{9}$Laboratori Nazionali di Frascati, INFN, I-00044 Frascati, Rome, Italy \\
$^{10}$Dept. of Physics, Univ. of Illinois at Urbana-Champaign, IL 61801, USA \\
$^{11}$Center for Exploration of Energy and Matter, Indiana Univ., Bloomington, IN 47408, USA \\
$^{12}$Istanbul Technical University, Istanbul 34469, Turkey \\
$^{13}$Royal Holloway, University of London, Egham, Surrey, UK \\
$^{14}$Dept. of Physics, University of Massachusetts, Amherst, MA 01003, USA \\
$^{15}$Dept. of Physics and Astron., Michigan State University, East Lansing, MI 48824, USA \\
$^{16}$Department of Physics, University of Patras, 26500 Rio-Patras, Greece \\
$^{17}$Physics Department, University and INFN Pisa, Italy \\
$^{18}$Dipt. di Fisica dell’Univ. di Roma ``Tor Vergata''
and INFN Sezione di Roma Tor Vergata, Rome, Italy \\
$^{19}$CEA/Saclay, DAPNIA, 91191 Gif-sur-Yvette Cedex, France \\
$^{20}$Dept. of Physics, University of Virginia, Charlottesville, VA 22904, USA \\
$^{21}$RWTH Aachen University and JARA-Fame, III. Physikalisches Institut B, Physikzentrum, 52056 Aachen, Germany \\
$^{22}$Harvard College, Harvard University, Cambridge, MA 02138, USA\\
$^{23}$Center for Axion and Precision Physics Research, IBS, Daejeon 305-701, Republic of Korea\\
$^{24}$Dept. of Physics, KAIST, Daejeon 305-701, Republic of Korea\\
$^{25}$TRIUMF, 4004 Wesbrook Mall, Vancouver, BC Canada V6T2A3\\
$^{26}$Fermi National Accelerator Laboratory, Batavia, Illinois 60510, USA\\
$^{27}$Dept. of Physics, University of Liverpool, Liverpool, UK\\
$^{28}$Korea Research Institute of Standards \& Science, Daejeon, Republic of Korea\\
$^{29}$Tech. Univ. M\"{u}nchen, Physikdepartment and Excellence-Cluster ``Universe'', Garching, Germany\\
$^{30}$Faculty of Applied Mathematics and Control Processes, Saint-Petersburg State University, Russia\\
$^{31}$Physics Department and INFN Section of Genoa, 16146 Genoa, Italy \\
$^{32}$Physics Department,  Korea University, Seoul 136-713, Republic of Korea\\
\\
$^\dagger$Deceased\\
$^*$Corresponding author, yannis@kaist.ac.kr
}
\renewcommand{\[}{\begin{equation}}
\renewcommand{\]}{\end{equation}}
\renewcommand{\[}{\begin{equation}}
\renewcommand{\]}{\end{equation}}
\renewcommand{\v}{{\bf v}}
\newcommand{\E}{{\bf E}}
\newcommand{\B}{{\bf B}}
\newcommand{\s}{{\bf s}}
\renewcommand{\b}{\boldsymbol\beta}
\renewcommand{\d}{{\bf d}}
\newcommand{\m}{{\boldsymbol\mu}}
\newcommand{\om}{\boldsymbol\omega}
\date{\today}
\begin{abstract}

A new experiment is described to detect a permanent electric dipole moment of the proton with a sensitivity of $10^{-29}e\cdot$cm by using polarized ``magic'' momentum $0.7$~GeV/c protons in an all-electric storage ring. Systematic errors relevant to the experiment are discussed and techniques to address them are presented. The measurement is sensitive to new physics beyond the Standard Model at the scale of 3000~TeV.

\end{abstract}
\maketitle

\section{Introduction}

One of the outstanding problems in contemporary elementary particle physics and cosmology is finding an explanation for the observed matter-antimatter asymmetry of our universe, known as baryogenesis. Within the framework of the Big Bang, it appears that a much greater degree of CP-violation than provided by the Standard Model of particle physics is required. That suggests the necessary existence of New Physics (NP), with large CP-violating interactions as a key ingredient in understanding the early universe. Identifying that NP source, central to our very existence, would be a major intellectual achievement.

In the search for laboratory manifestations of new CP-violating effects, electric dipole moments (EDMs) play a crucial role. 
A non-zero permanent particle electric dipole moment (EDM) separately violates parity (P) and time reversal symmetry (T)~\cite{landau}.  So, assuming CPT invariance, CP must also be violated. Since Standard Model EDM predictions are much smaller than current experimental sensitivities, an observation of any particle's EDM with today's technology would signal discovery of NP.   If of sufficient strength, such a source could provide a possible explanation for baryogenesis. Some theories, which suggest that an EDM may be within experimental reach, include supersymmetry (SUSY)~\cite{susyref}, left-right symmetry~\cite{lrsref}, and multi-Higgs scenarios~\cite{higgsref}.  Here, we explore the possibility of a storage ring search and study of the proton EDM ($d_p$) at the unprecedented level of $10^{-29} e \cdot {\rm cm}$, an advance by nearly 5 orders of magnitude beyond the current indirect bound of $|d_p| < 7.9\times 10^{-25} e \cdot {\rm cm}$ obtained using Hg atoms~\cite{moledm}.  Observing the EDM from different simple systems is necessary to identify the source of any NP~\cite{engel13}.

This dedicated direct proton EDM study at the level of   $10^{-29} e \cdot {\rm cm}$  is sensitive to a generic NP mass scale  $\Lambda_{\rm NP}$ with CP-violating phase $\phi_{\rm NP}$ roughly satisfying~\cite{dipole_moments} (3000~TeV/$\Lambda_{\rm NP})^2 \tan{(\phi_{\rm NP})} >1$. For a phase of the order of 45 degrees, the 3000~TeV NP scale is being probed, while for NP generically parametrized by a scale of order 1~TeV, a relative phase sensitivity as small as $\phi_{\rm NP} \approx 10^{-7}$  would be reached. Many specific examples of the outstanding probing power of a proton EDM study at the  $10^{-29} e \cdot {\rm cm}$ level exist. Here, we point out that it constrains the $\theta_{\rm QCD}$ parameter at $10^{-13}$, 3 orders of magnitude below the current neutron EDM bounds. A more timely illustration in the current LHC era is a potential CP-violating chiral phase induced by a loop induced Higgs to 2-photon coupling (the relative pseudoscalar to scalar amplitudes, a measure of potential Higgs CP-violation). Such an effective coupling would lead to fermion EDMs via quantum loops, making for an overall effect of 2-loop order. The proton EDM experimental program envisioned in this paper would be sensitive to a relative pseudoscalar coupling of order $10^{-3}$, about 2 orders of magnitude below current electron EDM constraints. It should be noted that EDMs of the electron, proton and neutron may actually represent our only practical access to the very small Higgs coupling to light (first generation) fermions, requiring of course the caveat that CP is violated at an observable level in the $H \gamma \gamma$  interaction.  

Searching for a non-zero proton EDM in a dedicated storage ring presents an experimental opportunity to improve the current sensitivity by more than three orders of magnitude compared to the current neutron EDM experimental limit~\cite{baker}. The method we describe is based on the frozen spin method and uses an all-electric lattice, directly measuring spin precession due to a non-zero EDM in an electric field. The Storage Ring EDM Collaboration has made significant progress in developing the experimental design for an all-electric EDM measurement. In this paper,  we describe the fundamental experimental techniques and the specifications of the all-electric storage ring. We also present the systematic errors and the methods developed to address them.

\section{Experimental Method}
The EDM and magnetic moment in terms of the rest frame spin $\s$ are $\d = (\eta e/ 2mc)\s,$ and $\m =(ge/2m)\s$, respectively, where these relations define $g$ and $\eta$; $G = (g-2)/2$ defined; and $m$ is the particle mass. At rest, the spin precession of a particle in electric and magnetic fields $\E$ and $\B$ is governed by:
\[\frac{d\s}{dt} = \m\times\B + \d\times \E.\]

For a particle with velocity $\b = \v/c$, relativistic factor $\gamma = \left(1 - v^2/c^2 \right)^{-1/2}$,  $\b\cdot\E = 0$ and $\b\cdot\B = 0$, the spin precesses relative to the momentum with angular velocity $\om_a + \om_e$~\cite{bmt,fukuyama,khriplovich}, where: 
\[\begin{aligned}\om_a& = \frac{e}{m}\left[ G\B  -\left(G - \frac{1}{\gamma^2 - 1}\right)\frac{\b\times\E}{c}\right]  \\\om_e&=  \frac{\eta e}{2 m}\left(\frac{\E}{c} + \b\times\B \right).\end{aligned}\]

Spin precession in a storage ring has been successfully used to establish a limit on a muon EDM~\cite{muonedm1,muonedm2}. For certain $\E$ and $\B$ fields, the $(g-2)$ precession of the particle $\om_a$ vanishes~\cite{far,orlov,yannis,nel}. Thus, aside from an EDM contribution, the spin is frozen along the momentum direction. With $\B = 0$ and the ``magic'' $\gamma =  \sqrt{1 + 1/G}$ chosen, $\om_a = 0$ and $\om_e = (\eta e/2mc)\E$. The parameters for this condition in the case of a proton are shown in Table \ref{proton}.  In an all-electric storage ring~\cite{mane1,mane2}, a radial electric field causes the spin to precess out of the storage plane linearly on the time-scale of the fill. Details of a storage ring proton EDM experiment are given in~\cite{protonEDMprop}.  For long-lived, polarized beams, the gain in sensitivity by using the frozen spin method over indirect 
methods~\cite{muonedm1,muonedm2} is several orders of magnitude~\cite{protonEDMprop}.

\begingroup
\squeezetable
\begin{table}[H]
\caption{\label{proton}``Magic'' proton parameters to cancel the $(g-2)$ precession in an all-electric ring.}
\begin{ruledtabular}
\begin{tabular}{ c c c c c }
 $G_p$~\cite{gproton} & $\gamma$ & $\beta$ & $p$ & $E$ \\ \hline
1.792847 & 1.248107 & 0.598379 & 0.7007 GeV/c & 1.171 GeV
\end{tabular}
\end{ruledtabular}
\end{table}
\endgroup

\begin{figure}[H]
\centering
\includegraphics[scale= 0.15]{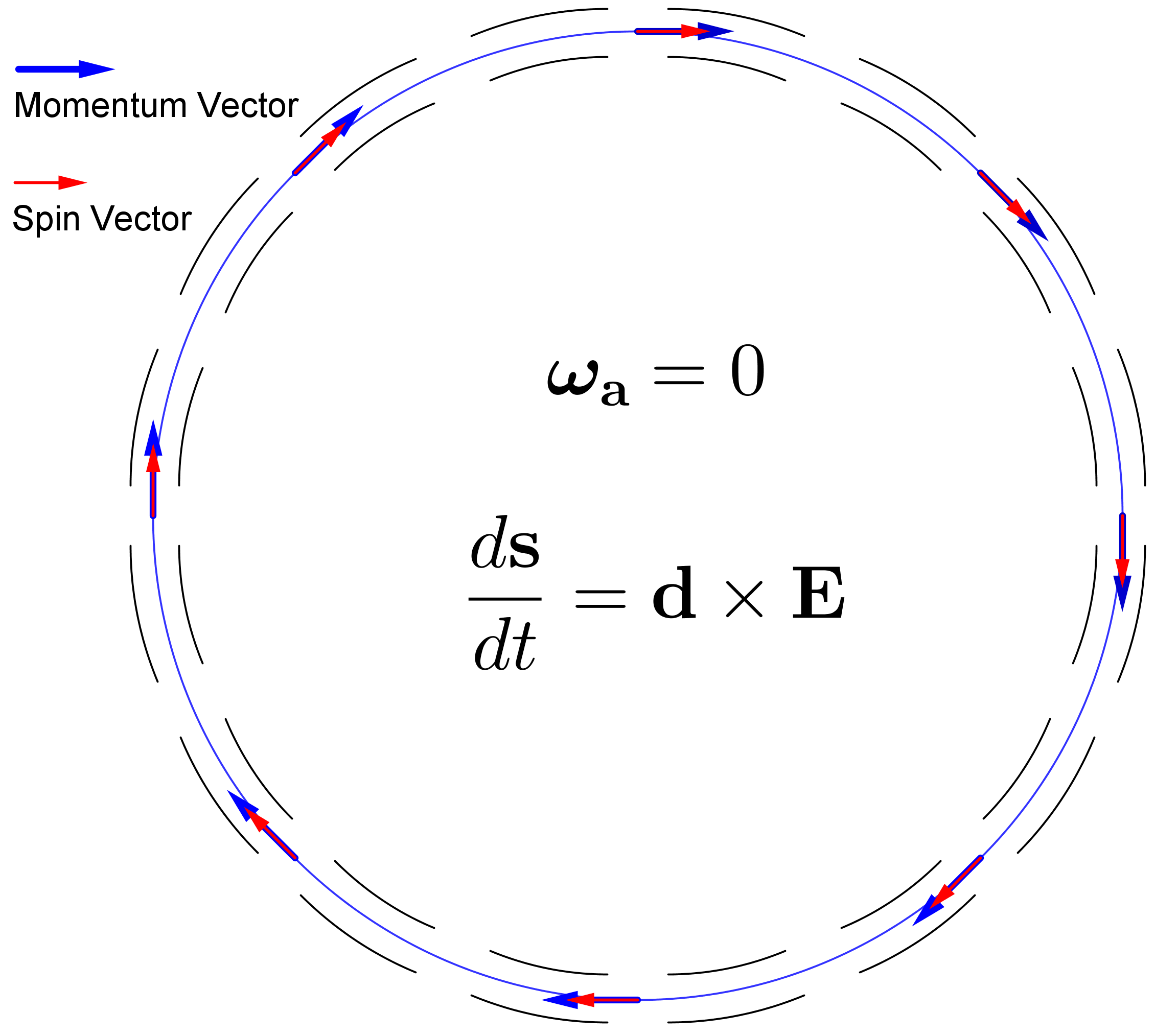}
\includegraphics[scale= 0.15]{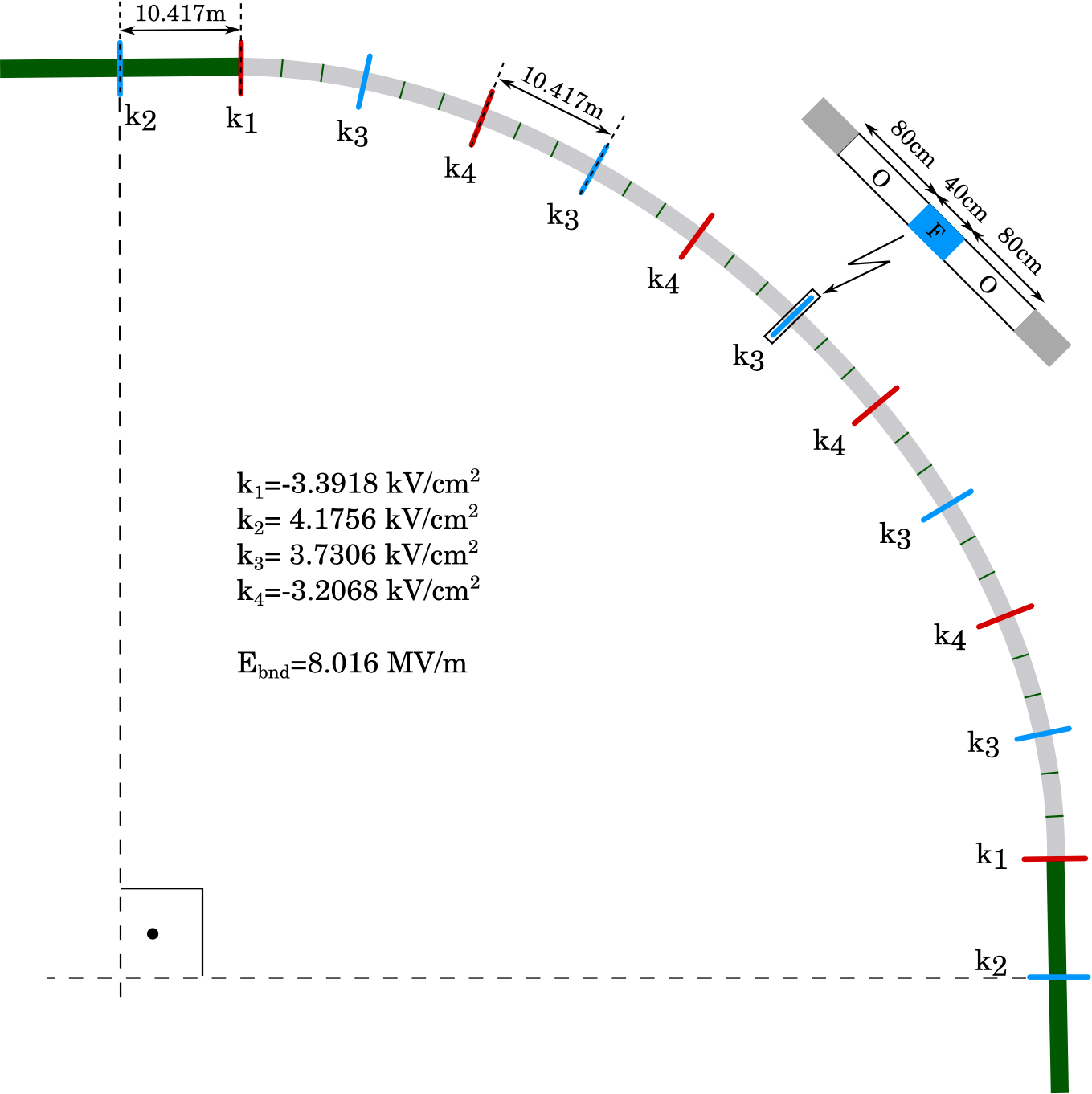}
\caption{\label{lattice}Top: An overview of an idealized ring with 16 cylindrical deflector segments and straight sections. The deflector bending radius is $R_0 = 52.3$~m and plate spacing is 3~cm. The electric field is directed radially inward between the plates. The spin and momentum vectors are kept aligned for the duration of storage. 
Bottom: A realistic lattice will include 40 bending sections separated by 36 straight sections 2.7~m long each, with electrostatic quadrupoles in an alternating gradient configuration, and four 20.8~m long straight sections for polarimetry and beam injection. It will also include SQUID-based magnetometers distributed around the ring, whose total circumference is 500~m.
}
\end{figure}

{\it Ring and beam parameters.} 
The all-electric ring geometry will include 40 sections of concentric cylindrical deflectors of 52.3~m bending radius, with 36 straight sections of 2.7~m length and four straight sections of 20.8~m length, adding up to a 500~m circumference. The 2.7~m straight sections will include superconducting quantum interference devices (SQUIDS) as magnetometers and electrostatic alternating gradient quadrupoles, with two polarimeters placed in the longer straight sections. Injection of the beams in opposite directions around the ring will occur in the remaining two 20.8~m straight sections. The deflector electric field will be about $8$~MV/m radially inward in the 3~cm spacing between the deflector plates.  In principle it is possible to modify the shape of the deflector plates to include vertical focusing.  The vertical bending specifications are strict, on the micrometer scale, and are under study.  An electrostatic storage ring of this size would be more than ten times larger than any previous electrostatic ring~\cite{essr1,essr2,essr3}. A simplified deflector lattice is shown in Figure \ref{lattice}.

{\it Experimental techniques.} One hundred bunches of $2.5\times 10^{8}$ vertically polarized protons will be injected in the clockwise (CW) direction and a similar number in the counter-clockwise (CCW) direction, circulating simultaneously. The beams will be allowed to de-bunch and then re-bunch at the required frequency using a high-harmonic ($h=100$ in this case) RF system. Using an RF solenoid, the spins of the protons will be rotated in the longitudinal direction, producing protons of both positive and negative helicities.  The vertical polarization difference between early and late times will determine the average vertical spin precession rate. The considered ring and beam parameters are summarized in Table \ref{ringbeamparameters}.

The horizontal spin coherence time (SCT) of the stored beam is the time required for the RMS spread in spin angles to become one radian.  Causes of a finite SCT include horizontal and vertical oscillations as well as longitudinal (energy) oscillations.  An RF-cavity is required to keep the SCT much longer than a few ms due to momentum spread of the beam. The vertical component of the proton spin grows linearly with time, at a rate on the order of nanoradians per second for an EDM detectable by the experiment. In practice, the linear growth of the vertical spin is limited by the SCT of the stored beam. Protons will be stored on the order of $10^3$~s, yielding an early-to-late change of the vertical spin component on the order of microradians.  A vacuum of better than $10^{-10}$~Torr is required to keep the beam stored in the ring for $10^3$~s.


The electric field of 8~MV/m between the cylindrical deflectors is comparable to previous work which has achieved similar field strengths~\cite{efieldwork1,efieldwork2,efieldwork3}. The effect of the fringe electric fields of the cylindrical deflectors has been investigated both analytically and by precision particle tracking~\cite{fringefields}. The bending radius of the plates has been adjusted to account for additional deflection due to fringe electric fields in the straight sections.

Intrabeam scattering (IBS) effects, increasing the stored beam phase-space parameters, set the time scale of the fill. The necessity of having sufficiently small IBS results in a moderate beam current limit. The resulting space-charge tune shifts are small. The beam-beam scattering effects are smaller and do not present a problem. 

\begingroup
\squeezetable
\begin{table}[H]
\caption{\label{ringbeamparameters}Ring and beam parameters of the proton EDM experiment.  Proton beam parameters refer to each storage direction.}
\begin{ruledtabular}
\begin{tabular}{ l r }
Bending radius, $R_0$ & 52.3~m\\
Electrode spacing, $d$ & 3~cm\\
Electrode height & 20~cm\\
Deflector shape & cylindrical\\
Radial E-field, $E_0$ & 8~MV/m\\
Number of straight sections & 40\\
Straight section lengths & 2.7389~m, 20.834~m\\
Polarimeter sections & 2\\
Injection sections & 2\\
SQUID-based magnetometer sections & 36\\
Total circumference, $C$ & 500~m\\
Harmonic number $h$, RF frequency & $100$, $35.878$~MHz\\
RF voltage, synchrotron tune $Q_s$ & 6~kV, 0.0066 \\
Particles per bunch & 2.5 $\times 10^8$\\
Maximum momentum spread, $(dp/p)_{\text{max}}$ & $4.6 \times 10^{-4}$\\
Horizontal beta function, $\beta_{x,\text{max}}$ & 47~m\\
Vertical beta function, $\beta_{y,\text{max}}$ & 216~m\\
Horizontal dispersion function, $D_{x,\text{max}}$ & 29.5~m\\
Horizontal tune, $Q_x$ & 2.42\\
Vertical tune, $Q_y$ & 0.44\\
Vertical emittance,  $\epsilon_{\rm Vmax}$ & 17~mm~mrad \\
Horizontal emittance, $\epsilon_{\rm Hmax}$ & 3.2~mm~mrad \\
Slip-factor, $\eta = \alpha - 1/\gamma^2$  & -0.192 
\end{tabular}
\end{ruledtabular}
\end{table}
\endgroup

\begin{figure}[H]
\centering
\includegraphics[scale = 0.45]{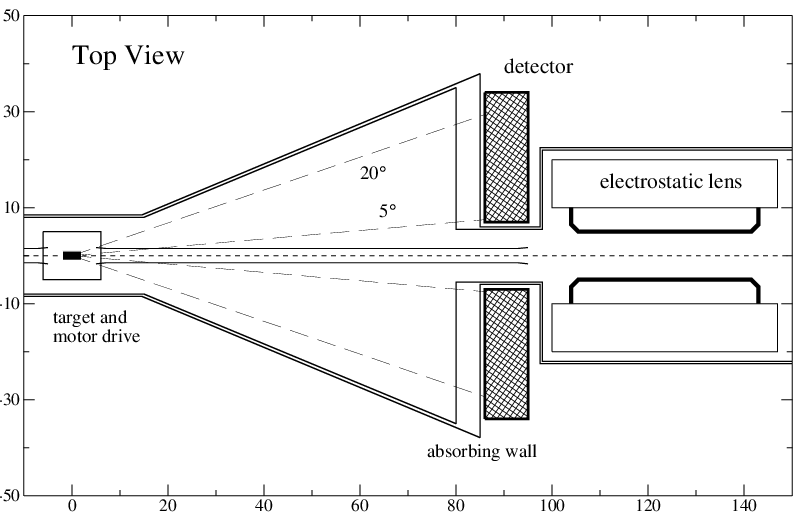}
\caption{\label{polarimeter}A possible layout of one-half of a proton EDM ring polarimeter, with the scale in centimeters. The carbon extraction target is located on a motor drive. The beam is denoted by a short-dashed line.}
\end{figure}

Polarimeters will be used to measure small changes in the vertical component of the beam polarization by scattering particles from a 6~cm thick carbon target.  The stored beam will be led to collide with the target using a number of possible alternative methods, e.g., by slowly lowering the vertical focusing strength, using a resonant slow extraction vertically, etc.  About 99\% of the time the beam particles undergo Coulomb scattering, lose enough energy and leave the ring.  Roughly 1\% of the time the protons undergo spin-dependent nuclear elastic scattering, ending up on a detector located about a meter beyond the target.
 The scattering of the particles in the up and down (left and right) directions will provide information on the horizontal (vertical) plane polarization component. Detectors must be able to respond to charged-particle events with minimal dead time and small systematic errors. Types under consideration include multi-resistive plate chambers, micro-megas chambers, gas electron multiplier chambers, and silicon detectors. A polarimeter design under consideration is shown in Figure \ref{polarimeter}.

A precision tracking program using numerical integration methods such as fourth-order Runge-Kutta and the Predictor-Corrector method~\cite{hamming59} is used to integrate the beam and spin dynamics differential equations. The Runge-Kutta and Predictor-Corrector integration methods used in tracking are slow but accurate; they are kept at a step size on the order of 1-10~ps.  The programs have been benchmarked to high accuracy against analytical estimates. Additional programs, such as UAL/ETEAPOT~\cite{ual,teapot} or using the Hamiltonian approach~\cite{mane14}, which account for the momentum change of the particle due to motion in E-fields, have been developed. Results from tracking simulations~\cite{selcuk,mane15} confirm that a particle distribution with $(dp/p) = 2\times 10^{-4}$ will have a SCT of the order $10^3$~s in an idealized lattice. The SCT of the stored beam was found to increase with increasing ring radius. 

For a uniform beam extraction rate, the statistical error $\sigma_{d}$ of the measurement has been both calculated analytically and confirmed numerically with Monte Carlo simulations to be:
\[\sigma_{d} = \frac{2\hbar}{P A E_0 \sqrt{N_{\text{tot,c}} T_{\text{tot}} f \tau_{\text{SCT}}}}.\]

Taking parameter values~\cite{Brantjes} of the beam polarization $P = 0.8$, the analyzer power $A = 0.6$, $E_0 = 8$~MV/m over 65\% of the ring, $N_{\text{tot,c}} = 5\times 10^{10}$ particles per storage cycle, a particle detection efficiency of $f = 0.0055$, a total running time $T_{\text{tot}} = 10^7$s per year, and a SCT of $\tau_{\text{SCT}} = 10^3$s, the statistical error for one year is $4\times 10^{-29}e\cdot$cm. The polarimeter analyzing power peaks near the proton ``magic'' momentum, a fortuitous coincidence.  The statistical error grows in proportion to the storage ring radius.  That error will be reduced by modulating the data-taking rate, taking most of the data at early and late times.  By further optimizing the ring lattice, reducing IBS and increasing SCT, we expect to be able to achieve a statistical sensitivity of $<2\times 10^{-29}e\cdot$cm per year.

\section{Systematic Errors}

Analytical estimates in combination with precision tracking allow the size of potential systematic errors to be estimated in several ways. Magnetic shielding, beam position monitors (SQUID-based BPMs sensitive to B-fields, plus button-BPMs sensitive to E-fields), and lattice alignment to better than 0.1~mm around the ring are sufficient to address the main systematic errors, i.e., radial B-fields and geometric phases~\cite{berry}. 

{\it Radial B-fields.} The ring will be shielded from the Earth's magnetic field as well as from noise through passive shielding and feedback mechanisms. The presence of a net radial B-field, $\left< B_r \right>$, could mimic an EDM signal, producing a vertical spin precession
\[\omega_V = \frac{e g \left< B_r \right>}{2 m \gamma^2}.\]
 An average radial field of  $10$~aT will cause a spin precession at the sensitivity of the EDM experiment. However, the radial magnetic field would split the CW and CCW beams vertically. The vertical beam position $\delta y$ as a function of the modes of the radial B-field is:
\[\label{deltay}\delta y = \sum_{N = 0}^\infty \frac{\beta c R_0 B_{r,N}}{E_0(Q_y^2 - N^2)}\cos(N\theta + \varphi_N),\]
where $Q_y$ is the vertical tune, which will oscillate between 0.44 and 0.48 at a frequency on the order of $10$~kHz.

As noted above, the counter-rotating beams will be split vertically, with maximum separation equal to twice that given by Equation \ref{deltay}. For a  $10$~aT magnetic field, the split between the beams is on the order of a  picometer. Beam position monitors (BPMs) are required that can determine the vertical positions of the CW and CCW beams with picometer-scale resolution over the $10^7$~s duration of the experiment. SQUID magnetometers are suitable to measure the magnetic field resulting from the splitting of the CW and CCW beams. Low-temperature DC SQUIDs have demonstrated sensitivities down to 1~fT/$\sqrt{\text{Hz}}$~\cite{squid1,squid2}.  Measuring an average splitting between the CW and CCW beams on the picometer level is feasible with appropriate SQUID placement in the straight sections of the lattice. In principle, no magnetic shielding will be required with continuous BPM measurement around the ring. The finite number of detectors limits the magnetic field modes to which the system is sensitive by the Nyquist sampling theorem.  The critical parameter is the detected average radial B-field and that can be wrong only when the mode is equal to the number of the BPM locations around the ring as well as its integer multiples. However, it has been shown analytically as well as by beam/spin tracking simulations that the SQUID-based BPMs are quite insensitive to higher harmonics of the radial B-fields.  SQUIDS are sensitive to time dependent B-fields only.  The ratio of the time oscillating component of $N=0$ to higher harmonics can be shown to be $(Q_y / N)^4$, making the high $N$~harmonic contributions negligible.
The shielding requirements are dominated by the so-called geometric phase effect (see below).  For this reason, the total magnetic field will be shielded to below 10-100~nT at all points in the ring.  Such shielding of the B-field is within the present state of the art~\cite{ptb}.

\begin{figure}[H]
\centering
\includegraphics[scale = 0.4]{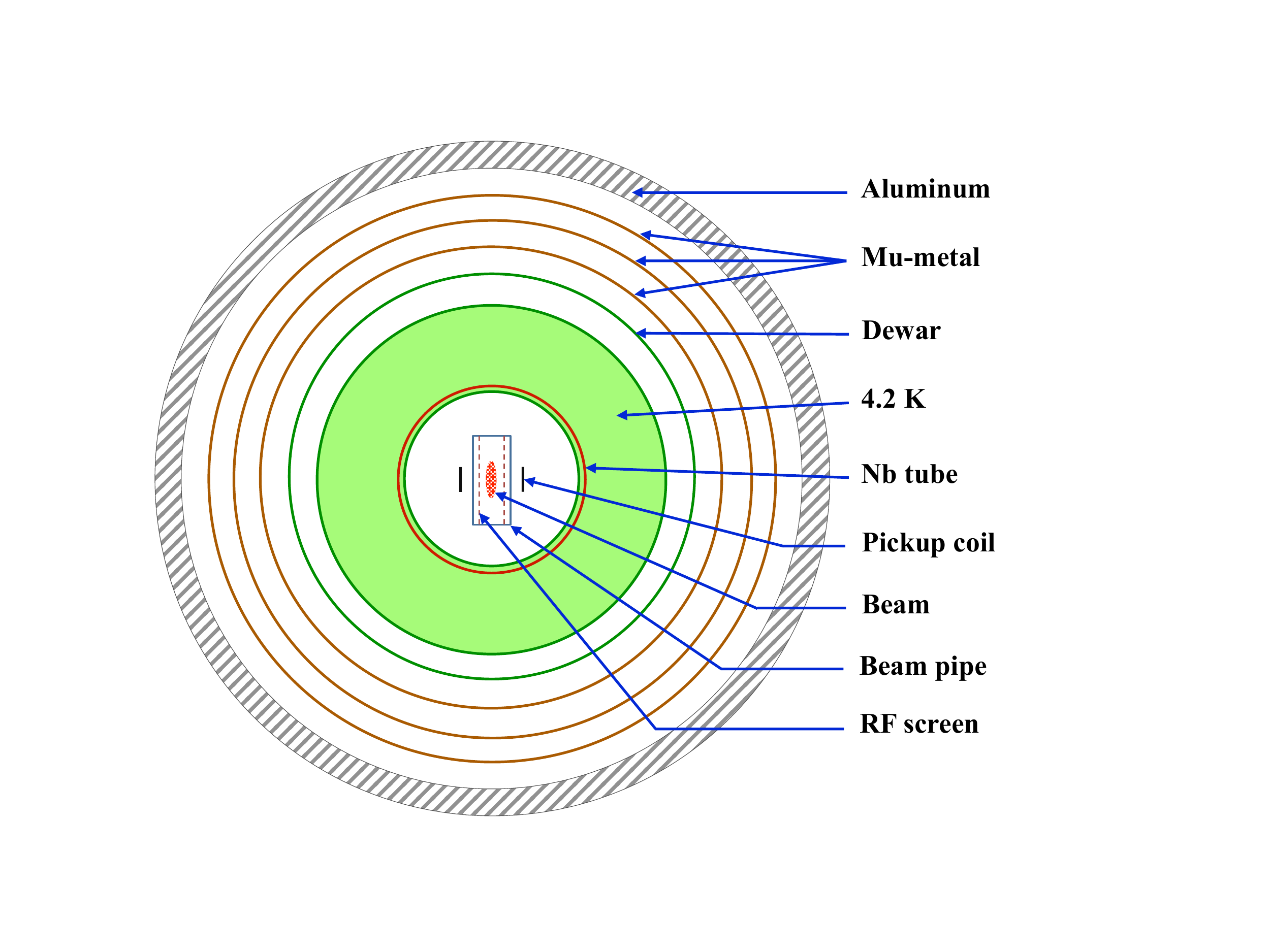}
\caption{\label{squid}A schematic of a possible SQUID BPM station. The system is shielded with a superconducting Nb tube, Al tube for RF-shield,  and several mu-metal layers.}
\end{figure}

SQUIDs in the straight-sections of the ring should be sufficient to measure the vertical splitting. A schematic of one such possible SQUID BPM station is shown in Figure \ref{squid}.  The vertical spin precession rate as a function of the detected radial B-field will be plotted, with the EDM signal corresponding to the DC offset in the vertical axis.

{\it Geometric phases.} In three dimensions, spin rotations about different axes do not commute. This fact contributes to geometric phase-induced false EDMs, a significant systematic error in neutron EDM experiments~\cite{baker,berry,neutronedm1}.  In a storage ring, geometric phases of spin dynamics may be cancelled more simply than in a neutron trap.  Consider, for example, the $N$-modes of two non-commuting perturbations: spin rotation frequency around the vertical axis, $( \delta \omega_V)_N \cos{(N \omega_c t)}$, and spin rotation frequency around the longitudinal axis, $( \delta \omega_L)_N \cos{(N \omega_c t + \phi )}$.  The presence of these perturbations in the lattice leads \cite{orlov02} to a spin rotation around the radial axis with a frequency $\Omega_R = |( \delta \omega_V)_N ( \delta \omega_L)_N \sin{\phi}/(2 N \omega_c) |$, thus imitating an EDM signal. It follows from this formula that in order to cancel the geometric phase $\Omega_R t$, we only need to find, experimentally, a counter-perturbation for either of the two $N$-modes of the perturbed fields.  The same approach can be used for other modes.  

Perturbed E and B-fields induce distortions of the closed orbits, the detection of which will be used to check that the errors are being kept at an acceptable level~\cite{protonEDMprop}. With respect to achieving this level, analytic investigation has shown that the geometric effect of B-fields splitting the counter-rotating beams can be kept below the experimental sensitivity if the maximum B-field is kept below the 10-100~nT level everywhere; that E-field errors due to plate misalignments displacing both counter-rotating beams from their ideal location can be addressed by placing BPMs within 0.1~mm of the ideal orbit; and that small changes to deflector geometry will adequately compensate for systematic errors due to deflector fringe fields~\cite{fringefields}.   

{\it Polarimetry.}  An experiment investigating the management of geometric and rate-induced systematic errors in polarimetry was conducted at COSY-J\"ulich~\cite{cosy2} with a 1.7~cm carbon block target. Large systematic errors consisting of position and angle changes to the beam were made deliberately in order to generate easily measurable effects. At the level of geometric and rate errors expected for the proton EDM experiment, the results of the COSY-J\"ulich study indicate how to reduce systematic uncertainties in polarimetry to well below the level of sensitivity.  Positive and negative helicity bunches will be stored in the same direction, and a combination of observables will be used to identify systematic errors due to non-linearities~\cite{Brantjes}.

{\it Vertical forces.} The interaction of the CW and CCW beams may lead to a systematic error: if the two counter-rotating beams do not overlap completely, on average they will feel a vertical force from one another. The problem can be addressed as long as the SQUID BPMs are sensitive to the beam separation size, and feedback can be used to eliminate the signal. Any forces on the beams due to image charges on the top and bottom of the vacuum chamber will be minimized by using vertical metallic plates for almost the entire azimuthal extent of the ring. Results from numerical simulations indicate that the aspect ratio of the quadrupole plates can be chosen to reduce the effect when the counter-rotating beam intensities do not cancel exactly. Sextupole electric fields combined with different CW and CCW beam sizes can create a vertical splitting of the counter-rotating beams, setting the specifications for both.  Introducing short runs in between regular-length runs will address systematic errors related to the vertical spin component of the beams being correlated to the protons' phase space parameters.  

{\it Other effects.} There are potential systematic errors due to the gravitational field and rotation of the Earth.   For example, there is a false-EDM signal due to the vertical E-fields being balanced by the force of gravity at our level of sensitivity. Taking the difference between signals of the CW and CCW beams will cancel this effect.  Also, Coriolis and Sagnac~\cite{sagnac} effects due to the rotation of the Earth have been found to be below the experimental sensitivity. The RF cavity will account for the slightly different travel times of the CW and CCW beams around the ring by equalizing the frequencies. However, the Sagnac effect may place an upper limit (more than $10^3$~s) on how long counter-rotating beams will be stored with longitudinal polarization. 

Spin resonances may also contribute to a false EDM signal, although this contribution is decreased in the frozen spin ring by a factor $1 / \tau_{\rm SCT}$.  In addition, spin and beam resonances coincide in the frozen spin ring, so they will be dealt with together  (at a later stage of the project).

A summary of the main systematic errors in the experiment is given in Table \ref{syserr}.

\begingroup
\squeezetable
\begin{table}[H]
\caption{\label{syserr}Main systematic errors of the experiment and their remediation.}
\begin{ruledtabular}
\begin{tabular}{ l  | p{5cm} }
Effect & Remediation \\ \hline
Radial B-field & SQUID BPMs with 1~fT/$\sqrt{\text{Hz}}$ sensitivity eliminate it.\\
Geometric phase & Plate alignment to better than 100~$\mu$m, plus CW and CCW storage.  Reducing B-field everywhere to below 10-100~nT. BPM to 100~$\mu$m to control the effect.\\
Non-Radial E-field & CW and CCW beams cancel the effect.\\
Vert. Quad misalignment & BPM measurement sensitive to vertical beam oscillation common to CW and CCW beams.\\
Polarimetry & Using positive and negative helicity protons in both the CW and CCW directions cancels the errors.\\
Image charges & Using vertical metallic plates except in the quad region. Quad plates' aspect ratio reduces the effect.\\
RF cavity misalignment & Limiting longitudinal impedance to $10$k$\Omega$ to control the effect of a vertical angular misalignment. CW and CCW beams cancel the effect of a vertically misplaced cavity.
\end{tabular}
\end{ruledtabular}
\end{table}
\endgroup

\section{Conclusions}
The Storage Ring EDM Collaboration has designed an experiment using the frozen spin method and a dedicated storage ring to measure the proton EDM with an unprecedented sensitivity of $10^{-29}e\cdot$cm. We are currently developing prototypes to optimize the critical systems of the experiment, which include magnetic shielding, SQUID-based BPMs, polarimeter and electric field plates.  In parallel we are developing software for high precision and high efficiency spin and beam dynamics tracking.  The proton EDM measurement will provide a valuable probe of new physics beyond the Standard Model.

\section{Acknowledgements}
We wish to acknowledge support for the polarimetry and spin coherence time measurements at COSY-J\"ulich by Forschungszentrum J\"ulich.   IBS-Korea (project system code: IBS-R017-D1-2014-a00) partially supported this project.  DOE partially supported this project under BNL Contract No. DE-SC0012704.

\end{document}